\documentclass[conference]{IEEEtran}
\IEEEoverridecommandlockouts

\usepackage[backend=biber,style=ieee]{biblatex}
\usepackage{amsmath,amssymb}
\usepackage{booktabs}
\usepackage{graphicx}
\usepackage{listings}
\usepackage{xcolor}
\usepackage{url}
\usepackage[hidelinks]{hyperref}
\usepackage{enumitem}
\usepackage{tikz}
\usetikzlibrary{arrows.meta, positioning, fit, backgrounds,calc}
\usepackage{xspace}
\usepackage[T1]{fontenc}

\lstdefinestyle{compactcode}{
  basicstyle=\ttfamily\scriptsize,
  columns=fullflexible,
  keepspaces=true,
  breaklines=true,
  breakatwhitespace=false,
  showstringspaces=false,
  frame=single,
  framerule=0.2pt,
  xleftmargin=0pt,
  xrightmargin=0pt,
  aboveskip=0.35\baselineskip,
  belowskip=0.35\baselineskip,
  tabsize=2
}
\lstdefinestyle{dbcode}{
   basicstyle=\ttfamily\footnotesize,
   columns=fullflexible,
   breaklines=true,
   breakatwhitespace=false,
   frame=single,
   framesep=3pt,
   xleftmargin=3pt,
   xrightmargin=3pt,
   captionpos=b,
   showstringspaces=false
}

\pgfdeclarelayer{background}
\pgfdeclarelayer{foreground}
\pgfsetlayers{background,main,foreground}

\usepackage{tcolorbox}
\definecolor{amber}{rgb}{1.0, 0.49, 0.0}
\definecolor{beaublue}{rgb}{0.74, 0.83, 0.9}
\definecolor{brown}{rgb}{0.65, 0.16, 0.16}
\definecolor{caribbeangreen}{rgb}{0.0, 0.8, 0.6}
\definecolor{carmine}{rgb}{0.59, 0.0, 0.09}
\definecolor{celadon}{rgb}{0.67, 0.88, 0.69}
\definecolor{champagne}{rgb}{0.97, 0.91, 0.81}
\definecolor{classicrose}{rgb}{0.98, 0.8, 0.91}
\definecolor{corn}{rgb}{0.98, 0.93, 0.36}
\definecolor{cornflowerblue}{rgb}{0.39, 0.58, 0.93}
\definecolor{crimson}{rgb}{0.83, 0.0, 0.25}
\definecolor{darkblue}{rgb}{0,0,0.5}
\definecolor{darkgreen}{rgb}{0,0.5,0}
\definecolor{darkelectricblue}{rgb}{0.33, 0.41, 0.47}
\definecolor{purple}{rgb}{0.57,0.63,0.81}

\newif\iffinal
\finalfalse

\iffinal
  \newcommand\guidance[1]{}
  \newcommand\todo[1]{}
  \newcommand\katie[1]{}
  \newcommand\patrick[1]{}
\else
  \usepackage[colorinlistoftodos,prependcaption]{todonotes}
  \newcommand\guidance[1]{\todo[inline]{#1}}
  \newcommand\katie[1]{\todo[inline,color=cornflowerblue]{Katie: #1}}
  \newcommand\patrick[1]{{\todo[inline,fancyline,color=classicrose]{Patrick: #1}}}

\fi

\begin{document}

\title{Domain Bounds as a Silent-Fault Detector for AI-Ready Scientific Data}

\author{
\IEEEauthorblockN{Kathryn Knight}
\IEEEauthorblockA{Oak Ridge National Laboratory\\
Oak Ridge, Tennessee, USA\\
0000-0003-2976-0049}
\and
\IEEEauthorblockN{Patrick Widener}
\IEEEauthorblockA{Oak Ridge National Laboratory\\
Oak Ridge, Tennessee, USA\\
0000-0002-5882-0816}
\and
\IEEEauthorblockN{Heather Bort}
\IEEEauthorblockA{Oak Ridge National Laboratory\\
Oak Ridge, Tennessee, USA\\
0009-0008-8621-0215}
}

\maketitle

\begin{abstract}
We address an incipient type of data-based faults, driven by increasing amounts of data reuse in AI-enhanced computational science workflows. A dataset is created under specific conditions; its categories, measures, and labels depend on that context, but that context is usually unavailable to downstream users and is completely abandoned when datasets are used to train models. We propose the encoding and association of this context, called \emph{domain bounds}, with the data to which it applies. By specifying the conditions for valid reuse, a domain bound flags apparently valid data from being reused in an incompatible context, even when every value falls within its expected range. We present cases from computational science and biomedicine where missing domain bounds allow silent reuse errors, and show why description and provenance approaches do not detect them. Our detector uses standard data container technologies, catches faults drawn from a taxonomy of published cases, irrespective of dataset size. Against injected faults it caught 17 of 18 domain-bound reuses that lineage and descriptive records accepted.

\end{abstract}


\section{Introduction}
\label{sec:introduction}

Fault tolerance at scale has long distinguished between active and dormant faults. A hardware or otherwise active fault may announce itself through a signaled error, providing a detectable means to halt execution or intervene before it propagates. Dormant faults can introduce potentially undetected errors, which need not originate in hardware or software. The fault-tolerance/resilience research community has traditionally considered detection and mitigation strategies for active and dormant faults as a "computation-local" issue, encouraged by the dominance of single-program multiple-data/bulk-synchronous parallel computation~\cite{valiant1990} in leadership-scale modeling/simulation. These strategies (e.g., reasonableness bounds) assume the evidence needed to detect a fault is available within the computation itself.

The growing popularity of distributed computational workflows in high performance computing (HPC) contexts, as well as the increasing adoption of AI/ML components as participants in such workflows, has introduced another dimension to the traditional fault model. Such workflows communicate internally and externally through the data they ingest, manipulate, and create, as well as through data they reuse from other sources. Increasingly in scientific workflows, errors can propagate across component boundaries through data \cite{hiller2004, duan2019}.

Data need not be corrupted (as happens with single-node silent data corruption, for example) for error propagation to occur. We observe that scientific workflow errors may arise simply from \emph{reuse}: when a dataset is prepared under one set of conditions and later used under another. Such conditions may be baldly obvious or vanishingly obscure; data labels and associated concepts may be tied to a particular population, operational definition, measurement regime, or downstream task. When conditions at the point of data reuse differ from those obtained during data creation, the pipeline may still run to completion: files parse, schemas match, provenance records validate, and model training or inference completes without error. In this way, results can be both computationally valid and scientifically unwarranted\textemdash{}a silent failure. 

The problem is compounded by the use of AI models. A workflow without inference components might produce, consume, and reuse datasets, but those datasets could still be made available for post hoc inspection and comparison. Model training, however, renders the training data opaque, especially in the case of models reused from an external source. Reusing an externally created model accepts the conditions its training data was prepared under, without being able to see what they were. In computational science, unlike general LLM-style text prediction, those conditions determine whether a result is meaningful.

There are many examples of this type of error. In materials science, the band gap of a material is a key determinant of its electrical behavior. Because direct measurement is often impractical, the values reported in materials science databases are typically computed rather than observed. But these computations require choosing an approximation method, and different methods will return different band gaps for the same material. Borlido et al. compared common approximations against experimental values for 472 materials and found even the best of them off by around 0.5\,eV on average, in a quantity whose values typically run from zero to a few eV~\cite{borlido2019}. Thus, two materials databases can report different band gaps for the same material and both be correct. A column named \texttt{band\_gap\_ev} does not record which approximation method produced it, so merging two such columns derived from different databases may yield a table that is well formed, correctly typed, \emph{and} misleading.

In plasma physics, disruption predictors trained on one tokamak often perform poorly on another. This has motivated work on cross-tokamak transfer learning where the target event is the same, but its observed precursors are machine-specific, reflecting differences in geometry, diagnostics, operating conditions, time scale, and disruption mechanisms~\cite{zheng2023}. This dependence on local conditions extends across scientific domains. Brewer et al. identify data preprocessing patterns common to AI data pipelines in climate, fusion, life sciences, and materials, but show that AI-readiness requires contextual understanding of domain-specific data constraints~\cite{brewer2026}. In both cases the procedure travels between settings, but the conditions it was applied under do not.

Disagreements over concept representation in data have been thoroughly examined in biomedicine. Throughout this paper we examine sepsis, which is widely documented in the
medical literature and is a useful example of this kind of disagreement. Sepsis-3~\cite{singer2016} is the current consensus definition of sepsis, but it does not say when sepsis begins. Cohen et al. built three sepsis cohorts with MIMIC-III, a public database of intensive care records widely used in medical machine learning research. They used three possible interpretations: sepsis onset is the time of organ failure, the time of suspected infection, or the earlier of the two. Each cohort applied one of them, and they disagreed about who counted as septic, giving cohorts of between 867 and 2,178 admissions from the same data. They then trained tree-based, deep learning, and survival models on each cohort, with inputs stopping at onset so the models forecast the condition rather than the labeling rule. Which Sepsis-3 interpretation produced the cohort mattered more to the result than which model was trained on it~\cite{cohen2024}. 

Weng et al. published a study using MIMIC-IV, MIMIC-III's successor, in which they identify septic patients by using ICD-9 and ICD-10 diagnosis codes~\cite{weng2025}. Wang and Zhang~\cite{wang2025} questioned that choice, as codes of that kind often miss cases that carry no coded diagnosis, and may admit patients with no physiologic evidence of infection or organ dysfunction. They recommend that later analyses apply the Sepsis-3 criteria, for which the MIMIC curators maintain a reference implementation, or report a sensitivity analysis. Cohen et al.'s result complicates that recommendation, as application of Sepsis-3 criteria admits more than one defensible reading. Thus, whether data can be reused depends on the interpretation that produced them, yet that interpretation is often recorded inconsistently or left unstructured. Reusers therefore inherit decisions they cannot inspect, and mismatches remain hidden until the data are interpreted differently.


This paper treats those unrecorded conditions as a fault class, and asks what it would take to detect a violation of these conditions before the resulting error can propagate. We call the conditions themselves \textit{domain bounds}. AI pipelines assemble training data by pooling sources, and no one inspects a corpus the way a reader inspects a paper. Highly contextual data may remain unquestioned as an AI/ML model training set, and a model fitted to such a set carries that context into every inference it serves.

We make the following contributions:
\begin{itemize}
    \item We present a fault model for domain-bound violations in scientific data reuse, with each class grounded in a published case documenting the corresponding interpretive mismatch. 
    \item We specify domain bounds as an authored record that compiles into an RO-Crate profile extension for exchange and into SHACL for validation, so that a consuming workflow can check a proposed reuse without a person in the loop. 
    \item We demonstrate how a machine-actionable domain bound fault detector can operate, through an experiment in which injected data reuse faults are subject to detection by provenance records, by descriptive metadata, and by domain bounds.
    \item We measure the consequences of undetected reuse, using two published sepsis cohorts drawn from the same population and selected by two definitions of the same condition.
\end{itemize}

The remainder of this paper is organized as follows. We discuss related work in Section~\ref{sec:related}. In Section~\ref{sec:fault}, we present the rationale for using domain bounds to detect data reuse errors and set out a fault taxonomy, drawn from cases in computational science and biomedicine, that we use to organize the discussion. Section~\ref{sec:mechanism} gives detail on one implementation of data reuse fault detection, using standard metadata description tools, and Section~\ref{sec:eval} describes our experimental validation, using data reuse faults derived from published research and our domain bounds specification mechanism. Section~\ref{sec:conclusion} presents concluding remarks.


\section{Background and Related Work}
\label{sec:related}
\subsection{Silent errors in HPC}

We use \textit{fault}, \textit{error}, and \textit{failure} as defined by Avizienis et al.~\cite{avizienis2004} and in the exascale resilience taxonomy of Snir et al.~\cite{snir2014}. A \textit{fault} is dormant until activated, an active fault causes an \textit{error} in the system's internal state, and the error becomes a \textit{failure} if it reaches the service interface. Snir et al. classify an error as detected when a message or signal indicates it, latent or silent when nothing does, and masked when it causes no failure. 

Algorithm-based fault tolerance dates back to Huang and Abraham, who encoded matrices with row and column checksums and showed that addition, multiplication, scalar product, LU decomposition, and transposition all preserve encoding, so a corrupted result could be checked against checksums the operations themselves preserved~\cite{huang1984}. Di and Cappello replaced fixed encoding with prediction, comparing each observed value against a tolerance range derived from a predicted next-step value, tuning the tolerance range to catch corruptions large enough to change a result~\cite{di2016, berrocal2015}. Each method requires some kind of reference for comparison: an encoding, a model, or a second copy of the computation.

Fiala et al. injected faults into MPI applications and found that a single error usually cascaded until it had reached every process, then replicated processes and compared the messages they sent in order to detect and correct it~\cite{fiala2012}. Benoit et al. addressed silent errors in linear workflows by combining detection with checkpointing at several levels~\cite{benoit2018}, and Duan et al. studied data resiliency in staging-based scientific workflows~\cite{duan2019}. A replica or a checkpoint supplies a reference, where a checksum or a predicted value supplied it before.

Silent errors, while understood at workflow scale, are currently treated as errors that depart from known concepts. We propose a further class, arising when a concept admits more than one defensible form, whether because it was revised over time or because several interpretations are in use at once. In this case, values are always correct but may disagree given certain contexts or timeframes.

\subsection{Semantic compatibility in scientific workflows}

Workflow researchers separate the structural type of a data object from what it means. Bowers and Ludäscher built a framework that kept the two apart and handled conversion between them~\cite{bowers2004}, later showing how semantic types could be propagated through a workflow~\cite{bowers2005}. The WINGS system reasoned over constraints attached both to components and to data sets, where the data set metadata recorded how a collection had been produced and preprocessed, and used those constraints to rule out workflow designs whose parts could not validly be combined~\cite{gil2011,kim2008}.

Belhajjame, Embury, and Paton classified the mismatches that arise when the output of one workflow operation is connected to the input of another~\cite{belhajjame2006}. They name four kinds: type, domain, representation, and extent, with cardinality a sub-case of type and content a sub-case of representation. This work is prior art for much of what a domain bound records, and several of these categories overlap with our fault classes. Of the four, extent is closest to the faults we describe: two parameters may agree on type, domain, and representation while drawing on non-overlapping sets of values, so a workflow that passes every structural check still cannot produce a result. Their example pairs two operations that both accept an open reading frame, a stretch of DNA that may encode a protein, one supplied from a Drosophila database and the other from a yeast database. 

The check they propose runs at composition time, against ontologies annotating each operation's parameters, so that two components can be connected. WINGS also validates before execution, and its constraints are held in component and data catalogs belonging to the system. A data set taken out of that environment retains none of this contextual information, as these constraints are part of system infrastructure and stay behind when data leaves.

We are concerned with an artifact already produced, whose conditions of production were fixed when it was made, and whose consumer may never touch the system that produced it. Composition-time checking has nothing to act on at that point, since the composition is long finished. A catalog has nothing to offer either, once the data has left the environment that maintained it. The reference has to travel with the artifact.

\subsection{Silent semantic failures in systems and AI pipelines}

Recent systems work detects failures where a program runs to completion while violating something it was supposed to satisfy. Oathkeeper infers semantic rules from the regression tests developers wrote for past failures, then enforces them at runtime~\cite{lou2022}. T2C derives checkers from existing test code directly~\cite{lou2025}. TrainCheck infers invariants for deep learning training and uses them to catch silent training errors~\cite{jiang2025}, and ETL-Compiler targets pipelines that are executable and schema-aligned while implementing incorrect logic~\cite{gupta2026}.

In each of these systems, checkers infer from a specification that predates the failure: a test asserting what a service should return, an API contract fixing the order of calls, a stated intent the pipeline was meant to implement. TrainCheck calls its targets objective correctness violations, arising from incorrect API usage, buggy library implementations, or faulty hardware~\cite{jiang2025}. 

Domain-bound faults admit no such specification. A cohort built under one reading of Sepsis-3 satisfies its own criteria, and so does a cohort built under another. Sepsis has no gold standard test, and so no way to confirm that definitions are being applied consistently across clinical settings or research studies~\cite{munroe2026}. A materials band gap computed under one approximation method is correct under that method, and so is a band gap computed under another. In neither case does a test separate the two, and nothing can be inferred from watching either behave correctly. The condition has to be declared rather than inferred.

In a scoped search of peer-reviewed HPC, distributed systems, scientific workflow, data management, and machine learning systems literature through August 2026, we found no treatment of this case as a detectable fault class for scientific data reuse.

\subsection{Limits of provenance, description, and shift detection}

A provenance record names the activities that an object has been subjected to and who or what was responsible for them. PROV-AGENT extends this to AI agents, recording their model invocations, prompts, responses, and telemetry in a single graph~\cite{souza2025}. Such a record can be complete and correct while the reuse it documents is invalid, since nothing in it states the conditions under which the object stops being valid.

Descriptive and FAIR-oriented metadata describe the artifact itself. Datasheets for Datasets asks whether anything about a dataset's collection or labeling might affect reuse, and whether there are tasks it should not be used for, with both answered in free text~\cite{gebru2021}. Gebru et al. state that the process is not intended to be automated. The first Dataset Nutrition Label was a set of JSON modules covering metadata, provenance, variables, statistics, and correlations~\cite{holland2018}. Its second generation asks a practitioner to select the intended use case for the model being trained, then shows the alerts specific to that use~\cite{chmielinski2020}. Chmielinski et al. report practitioner feedback that a single static label cannot serve every use of a dataset. The use cases are listed in advance by whoever built the label, and uses not on the list produce no alerts, which Chmielinski et al. name as an open problem. The label stays on the Data Nutrition Project website rather than traveling with the dataset, so a workflow still has nothing to validate against. 

Dataset shift detection compares a model's data inputs against the data it was trained on. Rabanser et al. tested a range of methods on simulated shifts and found that two-sample testing over the representations of a pre-trained classifier worked best, and that domain-discriminating classifiers can characterize a shift's type and whether it is harmful~\cite{rabanser2019}. Subbaswamy and Saria represent the data generating process as a causal graph, mark the components expected to shift between settings, and train models that are stable to those shifts~\cite{subbaswamy2020}. However, neither says which condition the reuse violated.

From our survey we conclude that a domain bound approach must satisfy three conditions: 1. it must supply a reference; 2. that reference must travel with the artifact rather than stay in the system that produced it; 3. it must be declared, since there is nothing to infer. 

\section{Using Domain Bounds to Characterize Data-reuse Faults}
\label{sec:fault}

A domain bound declares the conditions under which an artifact's categories were stabilized, and the word \emph{domain} complicates the meaning of that declaration. Ribes, writing on scientific infrastructure, treats domains as abstract conceptions, and shows that domains are fashioned according to individual perspectives and can be disputed afterward~\cite{ribes2019}. Albrechtsen argues that domains are constructed through the planning and design of knowledge organization systems rather than waiting to be discovered~\cite{albrechtsen2015}. A domain bound therefore records decisions that particular people made at a particular time. Artificial intelligence has a formal treatment of that dependence, using the term \emph{context}.

McCarthy's Turing Award lecture observes that for any axiom, there exists a broader context in which that axiom is not true as stated. No axiom escapes this, since there is no most general context to write it in~\cite{mccarthy1987}. He proposes that a sentence stored in a computer be read as true only in its named context rather than true everywhere, a relation he later formalized~\cite{mccarthy1993,mccarthy1998}. Guha's thesis takes up McCarthy's proposal and argues that knowledge bases do the opposite, storing each sentence as if it were true on its own, whatever else the base contains~\cite{guha1991}. Data repositories store values the same way, without the conditions under which they were produced. A domain bound records those conditions for a particular data artifact, and when they go unrecorded the artifact carries a domain-bound fault.

These dormant domain-bound faults can sit in a repository indefinitely. A reuse that is at odds with the artifact's particular conditions,   however, activates the domain-bound fault and produces an error: a cohort, a feature, or a model whose meaning differs from what the analyst takes it to be. The error then propagates, perhaps via a scientific workflow or model fine tuning, and manifests when an erroneous result is produced (a value, a prediction). Between activation and failure the error remains latent in the sense Snir et al. define, since the scientific pipeline continues to operate normally and reports nothing~\cite{snir2014}.

\subsection{Propagation and attribution}

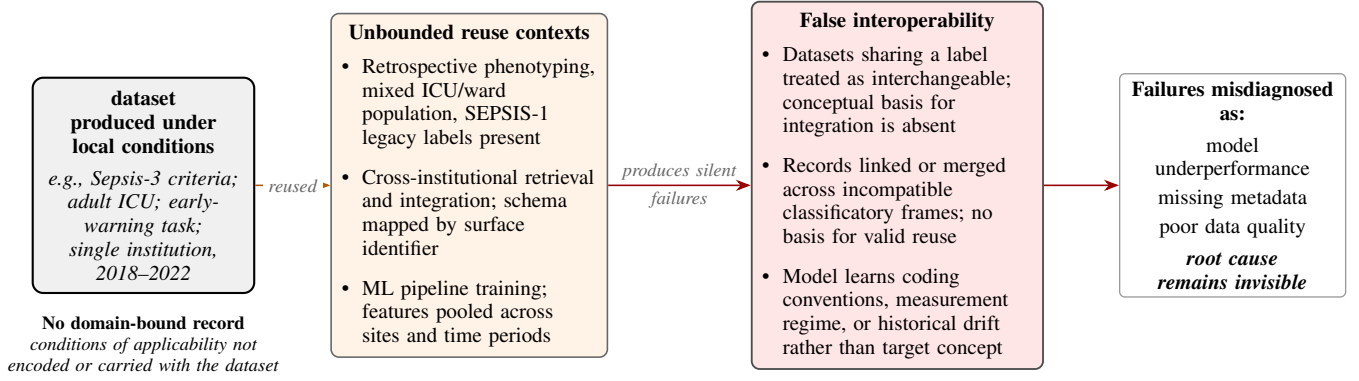
\begin{figure*}[t]
\centering
\resizebox{\textwidth}{!}{%
\begin{tikzpicture}[
  source/.style={
    rectangle, rounded corners=4pt,
    draw=black, line width=0.8pt,
    fill=gray!12,
    text width=3.2cm, align=center,
    font=\footnotesize, inner sep=5pt
  },
  context/.style={
    rectangle, rounded corners=3pt,
    draw=black!60, line width=0.6pt,
    fill=orange!10,
    text width=3.4cm, align=center,
    font=\footnotesize, inner sep=4pt
  },
  failure/.style={
    rectangle, rounded corners=3pt,
    draw=black!70, line width=0.7pt,
    fill=red!10,
    text width=4.0cm, align=center,
    font=\footnotesize, inner sep=4pt
  },
  annotation/.style={
    rectangle, rounded corners=2pt,
    draw=black!40, line width=0.5pt,
    fill=red!30,
    text width=4cm, align=center,
    font=\footnotesize, inner sep=3pt
  },
  mainarrow/.style={
    ->, >=Stealth, line width=0.7pt, draw=black!70
  },
  fanout/.style={
    ->, >=Stealth, line width=0.6pt, draw=orange!70!black
  },
  failarrow/.style={
    ->, >=Stealth, line width=0.6pt, draw=red!60!black
  },
  edgelabel/.style={
    font=\scriptsize\itshape, text=black!60,
    fill=white, inner sep=2pt
  },
  node distance=0.5cm and 1.1cm
]

\node[context, text width=3.4cm] (ctxbox) {
  \textbf{Unbounded reuse contexts}\\[4pt]
  \begin{itemize}[leftmargin=*, nosep, itemsep=6pt, label=\textbullet]
  \item Retrospective phenotyping, mixed ICU/ward population, SEPSIS-1 legacy labels present
  \item Cross-institutional retrieval and integration; schema mapped by surface identifier
  \item ML pipeline training; features pooled across sites and time periods
  \end{itemize}
};


\node[source, text width=2.6cm, left=1.0cm of ctxbox] (source) {
  \textbf{dataset produced under}\\
  \textbf{local conditions}\\[3pt]
  \textit{e.g., Sepsis-3 criteria;}\\
  \textit{adult ICU; early-warning task;}\\
  \textit{single institution, 2018--2022}
};

\node[font=\scriptsize, text width=4.0cm, align=center,
      below=5pt of source.south] (srccap) {
  \textbf{No domain-bound record}\\
  \textit{conditions of applicability not encoded or carried with the dataset}
};

\draw[fanout] (source.east) -- (ctxbox.west)
  node[edgelabel, midway] {reused};

\node[failure, text width=3.6cm, right=1.9cm of ctxbox] (failbox) {
  \textbf{False interoperability}\\[4pt]
  \begin{itemize}[leftmargin=*, nosep, itemsep=6pt, label=\textbullet]
  \item Datasets sharing a label treated as interchangeable; conceptual basis for integration is absent
  \item Records linked or merged across incompatible classificatory frames; no basis for valid reuse
  \item Model learns coding conventions, measurement regime, or historical drift rather than target concept
  \end{itemize}
};

\draw[failarrow] (ctxbox.east) -- (failbox.west)
  node[edgelabel, pos=0.5, above, fill=none] {produces silent}
  node[edgelabel, pos=0.5, below, fill=none] {failures};

\node[annotation, right=1.0cm of failbox,
      text width=2.8cm,text=black,fill=white] (consequence) {
  \textbf{Failures misdiagnosed}\\
  \textbf{as:}\\[3pt]
  model underperformance\\[2pt]
  missing metadata\\[2pt]
  poor data quality\\[3pt]
  \textit{\textbf{root cause \mbox{remains} invisible}}

};

\draw[failarrow] (failbox.east) -- (consequence.west);

\end{tikzpicture}
}%
\caption{When domain bounds are not carried forward with a dataset, downstream reuse can produce false interoperability. Datasets that share a label may be treated as interchangeable across retrieval, integration, and model-training contexts even when the conditions that made the label meaningful differ. Failures arising from this mismatch are frequently misdiagnosed as model underperformance, missing metadata, or poor data quality, obscuring the underlying problem of unbounded reuse.}

\label{fig:concept-drift-reuse}
\end{figure*}

Retrieval and integration systems need only a shared label to treat two datasets as interchangeable. Merging succeeds by matching strings or identifiers, even if the integrated datasets have labeled their records according to different rules or calculated values by different methods. We call this \emph{false interoperability}~(Figure~\ref{fig:concept-drift-reuse}): the data combine cleanly while the scientific warrant for combining them is weak or absent. 

In retrieval-augmented generation, false interoperability can occur within a single query. Records are selected by vector similarity, so two datasets that are semantically close are returned, and the generator writes output based on data from both. Nothing between the query and the output evaluates whether the retrieved records were produced according to compatible methods or domain conceptions.

Models are also trained across pooled data sources. Where the pooling includes sources from various locations or time periods, that can include data conceptualized according to local coding practice, workflow convention, measurement differences, and historical drift in the concept itself.

De Souza et al. describe agentic workflows in which agents invoke models and tools, and note that errors propagate when one agent's output becomes another's input~\cite{souza2025}. Where such an agent also selects datasets, there is no counterpart to a human-in-the-loop ensuring that domain bounds have been accounted for (i.e., there is no computational equivalent to Wang and Zhang's critical reading of sepsis cohort construction, as described in Section I).

Low model accuracy has many possible causes, as does poor transfer to a new site. For example, Subbaswamy and Saria pose a mortality prediction model that performs worse at a second hospital, where the drop in accuracy makes it difficult to say why, since shifts in patient demographics, in antibiotic prescribing habits, or in lab ordering patterns would all produce it~\cite{subbaswamy2020}. Snir et al. give HPC a vocabulary for the same problem, separating full diagnosis, which identifies the root cause of a failure, from partial diagnosis, which traces an error back through the causality chain without necessarily reaching the fault that started it~\cite{snir2014}. For a domain-bound fault, the chain runs back to a decision about a domain's chosen categorical constructs, taken once but left undocumented. Thus, only partial diagnosis is available, which stops at the dataset rather than the domain-bounded decisions and theories that shaped its construction.

\subsection{A taxonomy of injectable faults}

The taxonomy that follows is a device for showing where domain bounds differ from existing metadata practice, rather than a scheme we propose for adoption. We assembled it from cases already in the literature. In each case, two groups used the same data under different conditions and reported different results, and the reuse that produced the difference becomes a class. Every class therefore names something that has already gone wrong in print and can be checked against its source. The taxonomy is incomplete, since it can only contain what somebody has documented.

Table~\ref{tab:faults} lists the taxonomy classes, grouped by the type of metadata record that could catch each. The letter gives the group: D for domain-bound, L for lineage, M for descriptive metadata, R for residual. Lineage and descriptive classes are there so the experiment is not run only on faults our own method was built to catch. Residual classes are there because nothing in the present study catches them.

\begin{table}[t]
\caption{Fault classes, by the record positioned to detect them. Each
domain-bound class cites two sources, once in computational science and
once in biomedicine.}
\label{tab:faults}
\centering
\scriptsize
\begin{tabular}{llp{2.05cm}p{2.05cm}}
\toprule
ID & Class & Computational source & Biomedical source \\
\midrule
\multicolumn{4}{l}{\emph{Domain-bound}} \\
D1 & operational-definition drift & Borlido et al.~\cite{borlido2019} & Szakmany et al.~\cite{szakmany2018}; Wang and Zhang~\cite{wang2025} \\
D2 & onset reconstruction mismatch & Brewer et al.~\cite{brewer2026} & Cohen et al.~\cite{cohen2024} \\
D3 & population transfer & Zheng et al.~\cite{zheng2023} & Szakmany et al.~\cite{szakmany2018} \\
D4 & category conflation & Borlido et al.~\cite{borlido2019} & Vincent~\cite{vincent2023} \\
D5 & task mismatch & Brewer et al.~\cite{brewer2026} & Munroe and Prescott~\cite{munroe2026} \\
D6 & measurement regime & Zheng et al.~\cite{zheng2023} & Rhee et al.~\cite{rhee2017} \\
D7 & site or temporal pooling & Zheng et al.~\cite{zheng2023} & Subbaswamy and Saria~\cite{subbaswamy2020} \\
D8 & invalid transformation & Brewer et al.~\cite{brewer2026} & Rabanser et al.~\cite{rabanser2019} \\
D9 & missingness convention & Brewer et al.~\cite{brewer2026} & Rhee et al.~\cite{rhee2017} \\
\midrule
\multicolumn{4}{l}{\emph{Lineage}} \\
L1 & unregistered artifact & \multicolumn{2}{l}{Longpre et al.~\cite{longpre2024}} \\
L2 & missing lineage & \multicolumn{2}{l}{Souza et al.~\cite{souza2025}} \\
\midrule
\multicolumn{4}{l}{\emph{Descriptive}} \\
M1 & incomplete metadata & \multicolumn{2}{l}{Gebru et al.~\cite{gebru2021}} \\
M2 & format mismatch & \multicolumn{2}{l}{Hiniduma et al.~\cite{hiniduma2025}} \\
M3 & license unstated & \multicolumn{2}{l}{Longpre et al.~\cite{longpre2024}} \\
\midrule
\multicolumn{4}{l}{\emph{Residual}} \\
R1 & mis-authored bound & \multicolumn{2}{l}{Wang et al.~\cite{wangdrift2011}} \\
R2 & unmodeled dimension & \multicolumn{2}{l}{Lebovitz et al.~\cite{lebovitz2021}} \\
\bottomrule
\end{tabular}
\end{table}

In the band gap case from Section~\ref{sec:introduction}, a value computed by one approximation method, a value computed by another, and a measured value are three different quantities sharing a column name. Pooling them breaks the definition each was produced under and ignores that computed and measured values are not interchangeable, which is D1 and D4 together. Nothing about the files will show it. The tokamak case is D3, D6, and D7 together: the machine, the instruments, and the operating conditions all differ while the label stays the same. 

Brewer et al. survey data readiness pipelines in climate, fusion, life sciences, and materials. The same preprocessing steps recur across all four, but each field imposes constraints of its own~\cite{brewer2026}. In gridded climate data, an absent value may mean an unobserved cell, a masked region, or a physical zero. The data type does not say which, so a domain bound states the convention the file follows (D9). 

We use sepsis as the worked case for our evaluation, because its discrepancies have been quantified in print more thoroughly than those in the computational examples. Vincent argues that sepsis names a construct built from infection and organ dysfunction rather than a fixed disease, and that the criteria for identifying it have changed over time~\cite{vincent2022}. He later argues that infection and sepsis should not be used interchangeably, since infection alone is not enough for sepsis~\cite{vincent2023}.

Szakmany et al. applied the SEPSIS-1 and SEPSIS-3 criteria to the same 380 ward and emergency department patients across 13 hospitals. The first identified 212 septic patients, the second 272, and qSOFA 50~\cite{szakmany2018}. Munroe and Prescott note that sepsis definitions are applied in clinical care, research, and surveillance, and that these settings place different demands on sensitivity, specificity, timeliness, and ease of implementation~\cite{munroe2026}. Which definition is right therefore depends on what it is for.

%
%

\section{Mechanism}
\label{sec:mechanism}
This section describes how a domain bound is written, carried with the data it describes, and checked when a reuse is proposed. The record is authored once, by hand. Everything after that is derived from it automatically: a package that carries the bound wherever the data goes, and a set of constraints evaluated against any proposed reuse. Both use existing standards for attaching metadata to data, so a bound travels with the data rather than staying in the system that produced it.

\subsection{Authoring}

The authored record, a \texttt{DomainBoundSpec}, is a flat set of fields naming the conditions under which a dataset's categories, measures, and labels were stabilized, meaning fixed for a particular purpose at a particular time. Listing~\ref{lst:spec} shows an abbreviated but working record for the sepsis example from Section~\ref{sec:introduction}.

\begin{lstlisting}[style=dbcode, caption={An abbreviated authored record.}, label={lst:spec}]
{
  "recordType": "DomainBoundSpec",
  "profile": "sepsis_labeling",
  "artifact": "sepsis_labels.csv",
  "population": "adult",
  "setting": "icu",
  "siteScope": "single_institution",
  "temporalScope": "2008-2019",
  "operationalDefinition": "doi:10.1001/jama.2016.0287",
  "measurementRegime": "ehr_structured",
  "onsetReconstructionMethod": "retrospective_ehr",
  "intendedUse": "early_warning",
  "validTransform": ["sofa_scoring", "label_derivation"],
  "notInterchangeableWith": [
    "infection_label", "icd_coded_sepsis_label"
  ]
}
\end{lstlisting}

Fields fall into two categories. Interpretive fields, such as \texttt{intendedUse}, \texttt{notInterchangeableWith}, \texttt{portabilityLimits}, and \texttt{assumptions}, require knowledge belonging only to the creator of the dataset and cannot be filled in automatically. The rest can in principle be populated from information already available to the scheduler, orchestrator, or process environment: cohort definitions, data dictionaries, or workflow parameters. The sepsis record has values in 16 fields, of which 4 require an author and 12 could be derived. Authoring therefore means filling in four fields, not sixteen. Production data-reuse workflows often already record provenance and other metadata, which can supply some of the twelve.


Each record has a content digest, a truncated SHA-256 hash of its canonical serialization, which serves as its identifier. Bounds are revisable, and editing one changes its digest, so a revision appears under a new identifier rather than silently replacing the old record.

Each interpretive field states something settled when the data was constructed. In this instance, that is the population selected, the definition applied, the task the labels were fixed for, and the categories known at the time to be incompatible. Authoring one therefore does not require anticipating future uses. If a bound requires revision (say, if categories are deprecated or updated), the content digest will change with the updated bound, leaving earlier versions as-is.

\subsection{Exchange}

We organize the record according to the RO-Crate~\cite{rocrate2022} specification. A production implementation would extend the publicly available RO-Crate definitions through  established mechanisms (pull requests against the public RO-Crate repositories); here we define them locally to demonstrate feasibility and intend to contribute them as our work matures.

To conform to RO-Crate requirements, we add one profile, \texttt{DomainBoundedWR}, to the existing Workflow Run RO-Crate composition~\cite{leo2024workflowrun}. Workflow Run RO-Crate already composes:

\begin{itemize}
    \item Process Run Crate (a representation of an execution of a tool, recorded as a \texttt{CreateAction} with a \texttt{SoftwareApplication} or \texttt{ComputationalWorkflow} instrument);
    \item Workflow Run Crate (an explicit workflow definition referenced as \texttt{mainEntity} of the root \texttt{Dataset}); and 
    \item Workflow RO-Crate~\cite{workflowrocrate} (the description of the workflow itself). 
\end{itemize}
\texttt{DomainBoundedWR} extends this composition rather than replacing it, naming every profile in the chain on the root \texttt{Dataset} entity, as shown in Listing~\ref{lst:conformsto}.

\begin{lstlisting}[style=dbcode, caption={Root \texttt{Dataset} conformance chain in a \texttt{DomainBoundedWR} crate.}, label={lst:conformsto}]
"conformsTo": [
  {"@id": "https://w3id.org/ro/wfrun/process/0.5"},
  {"@id": "https://w3id.org/ro/wfrun/workflow/0.5"},
  {"@id": "https://w3id.org/workflowhub/workflow-ro-crate/1.0"},
  {"@id": "https://w3id.org/domain-bounds/wr/0.1"}
]
\end{lstlisting}

This is accompanied by a \texttt{mainEntity} reference to the \texttt{ComputationalWorkflow} entity, as Workflow Run Crate's own conformance rule requires. A data entity with a bound points at a \texttt{DomainBoundSpec} contextual entity through a \texttt{db:domainBounds} property, and the bound entity's identifier is the content digest described above. Because RO-Crate profile conformance is additive, a consumer that has never heard of \texttt{DomainBoundedWR} reads the crate as an ordinary Workflow Run Crate and disregards the extra entity, so the extension can be adopted without disturbing existing tooling.

\subsection{Validation}
SHACL (Shapes Constraint Language), expresses validation checks over linked-data containers such as RO-Crate. We use it because it is standardized and because engines are readily available for implementers~\cite{figuera2021,ahmetaj2025,ahmetaj2025_2}. What follows describes how a domain bound held in a crate is turned into constraint shapes and checked against a proposed reuse.


We continue working with the sepsis example. At reuse time a consuming workflow states its intent as a \texttt{ReuseRequest}, giving its task, population, setting, assumed operational definition, and planned transformations on the dataset. The bound is then compiled into a set of property constraints targeting that request, under three rules: 
\begin{itemize}
    \item a scope field, such as the operational definition, becomes a single-member \texttt{sh:in} constraint on the corresponding request field;
    \item the list of valid transforms becomes one multi-member \texttt{sh:in} constraint; and
    \item each declared non-interchangeable category becomes two negated \texttt{sh:not}/\texttt{sh:hasValue} constraints, one over the requested reuse category and one over the set of category the request proposes to pool the artifact with.
\end{itemize}

\texttt{sh:in} is used rather than a bare \texttt{sh:hasValue} because \texttt{sh:hasValue} requires the property to exist, whereas \texttt{sh:in} is trivially satisfied when it is absent.



Listing~\ref{lst:shacl} shows two of the resulting constraints: the one on the operational definition, and one of the pair generated by the ICD-coded label that the sepsis record declares itself not interchangeable with. 

The 16 populated fields of the sepsis record compile to 17 constraints. Ten come from the artifact reference and the nine scope fields, one from the transform list, and six from the three non-interchangeable categories. The remaining four fields, among them the free-text portability limits and assumptions, generate no constraint at all. The result is an ordinary SHACL Core~\cite{shacl2017} node shape targeting the class \texttt{db:ReuseRequest}, so a compiled bound can be checked by any conforming SHACL engine. 

\begin{lstlisting}[style=dbcode, caption={Two compiled reuse constraints, rendered as SHACL Core.}, label={lst:shacl}]
db:ReuseShape a sh:NodeShape ;
  sh:targetClass db:ReuseRequest ;
  sh:property [
    sh:path db:operationalDefinition ;
    sh:in ( "doi:10.1001/jama.2016.0287" ) ;
    sh:message "reuse assumes a different operational definition than the one the label was derived under" ;
  ] ;
  sh:not [ sh:property [
    sh:path db:reuseAs ;
    sh:hasValue "icd_coded_sepsis_label" ;
  ] ] .
\end{lstlisting}


A request that leaves a dimension unasserted (that is, presenting a simple scalar value) is treated as unconstrained on that dimension rather than a violation. A consuming workflow may know its population and task without knowing the measurement regime under which the artifact was produced, and requiring every dimension to be asserted would reject legitimate partial requests.

A blank field therefore never causes a violation. A violation names the condition that failed, so a workflow learns which of the dataset's stated conditions its proposed reuse does not meet. Validation can be completely automated at workflow orchestration or job submission time and does not require examination of the data in the dataset, so the cost of a check does not vary with the size of the dataset. 


\subsection{Generality and Limits}

Domain bounds do not depend on SHACL or RO-Crate. Both are convenient and well-known implementation choices, and what they provide is a bound that can be serialized, checked, and carried forward with the artifact. Any mechanism meeting those three requirements would serve.

The mechanism compares a stated reuse against an authored bound; it does not verify that the authored bound describes the dataset correctly. A bound that is present, well-formed, and wrong about the population or method it names is read and trusted like any other, and a reuse that is inconsistent with the true conditions of the artifact but consistent with the erroneous bound will not be caught. A hazard along a dimension the profile does not represent similarly leaves no constraint to violate. We discuss instances of both in the next section.

\section{Evaluation}
\label{sec:eval}


This section tests two claims. The first is that a domain bound can be checked programmatically, and that checking it catches reuse faults which provenance records and descriptive metadata do not. The second is that those faults have consequences worth catching. We describe the experimental setup, report what each detector caught, measure what one uncaught reuse costs using published patient data, and close with the fault classes nothing in this study catches.

\subsection{Setup}

\begin{figure}[t]
\centering
\resizebox{\columnwidth}{!}{
\begin{tikzpicture}[
  profile/.style={
    rectangle, rounded corners=3pt,
    draw=black!60, line width=0.6pt,
    fill=gray!12,
    text width=3.0cm, align=center,
    font=\footnotesize, inner sep=4pt
  },
  taxonomy/.style={
    rectangle, rounded corners=3pt,
    draw=black!60, line width=0.6pt,
    fill=orange!10,
    text width=3.0cm, align=center,
    font=\footnotesize, inner sep=4pt
  },
  inject/.style={
    rectangle, rounded corners=3pt,
    draw=red!60!black, line width=0.7pt,
    fill=red!6,
    text width=5.0cm, align=center,
    font=\footnotesize, inner sep=4pt
  },
  detector/.style={
    rectangle, rounded corners=3pt,
    draw=blue!50!black, line width=0.6pt,
    fill=blue!7,
    text width=3.1cm, align=center,
    font=\footnotesize, inner sep=4pt
  },
  outcome/.style={
    rectangle, rounded corners=3pt,
    draw=black!60, line width=0.7pt,
    fill=black!4,
    text width=6.4cm, align=center,
    font=\footnotesize, inner sep=5pt
  },
  arr/.style={-{Stealth[length=2mm]}, line width=0.6pt, draw=black!65},
  node distance=6mm and 6mm
]

\node[profile] (profile) {
  \textbf{Authored profile}\\
  crate + reuse request\\
  \textit{e.g.\ sepsis labeling}
};

\node[taxonomy, right=6mm of profile] (taxonomy) {
  \textbf{Fault taxonomy}\\
  16 classes, 4 groups\\
  D\,9 \quad L\,2 \quad M\,3 \quad R\,2\\
  + benign controls
};

\node[inject, below=18mm of $(profile)!0.5!(taxonomy)$] (inject) {
  \textbf{Fault injection}\\
  mutate the crate and/or the reuse request per fault class
};

\node[detector, below left=10mm and -2mm of inject] (prov) {
  \textbf{prov}\\
  reads lineage:\\
  registration, agent,\\
  inputs
};
\node[detector, below=10mm of inject] (fair) {
  \textbf{fair}\\
  reads descriptive\\
  metadata: type, format,\\
  license
};
\node[detector, below right=10mm and -2mm of inject] (bounds) {
  \textbf{bounds}\\
  reads domain bounds:\\
  population, definition,\\
  task, \ldots
};

\node[outcome, below=8mm of fair] (outcome) {
  \textbf{Per-trial outcome}\\
  each detector reports flagged or not flagged; outcomes are recorded per profile, fault class, and benign variant
};

\draw[arr] (profile.south) -- (inject.north);
\draw[arr] (taxonomy.south) -- (inject.north);

\draw[arr] (inject.south) -- ++(0,-4mm) -| (prov.north);
\draw[arr] (inject.south) -- (fair.north);
\draw[arr] (inject.south) -- ++(0,-4mm) -| (bounds.north);

\draw[arr] (prov.south) |- ($(prov.south)!0.5!(outcome.north west)$) -- (outcome.north west);
\draw[arr] (fair.south) -- (outcome.north);
\draw[arr] (bounds.south) |- ($(bounds.south)!0.5!(outcome.north east)$) -- (outcome.north east);

\end{tikzpicture}}
\caption{Structure of the fault-injection experiment. Sixteen fault classes across four groups, plus benign controls, mutate the authored crate and/or its reuse request. All three detectors examine the same trial, each restricted to a different record of the crate, and per-trial outcomes are recorded across every profile, fault class, and benign variant.}
\label{fig:evaluation}
\end{figure}
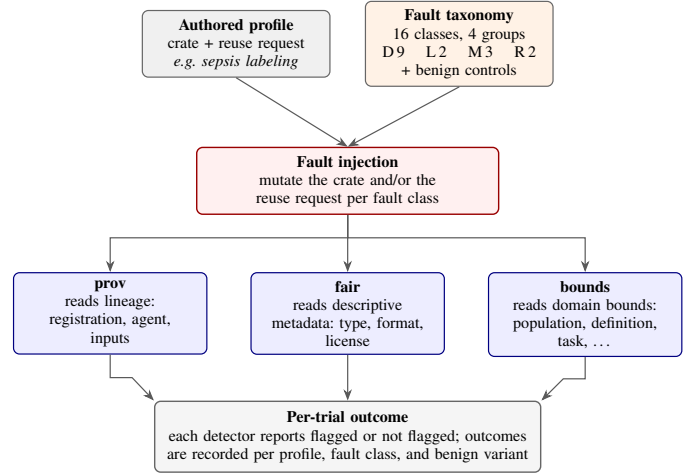

We ran the same experiment against two authored domain bound records, one describing materials science data and one describing biomedical data; the structure is shown in Figure~\ref{fig:evaluation}. Each record states the conditions under which the data categories were stabilized, and each is paired with a reuse request stating what a consumer proposes to do with those data. Neither the records nor the requests refer to the data itself. Our code is available at \url{https://github.com/keknight/domain-bounds-release}.

%

The sepsis domain bound record is shown in Listing~\ref{lst:spec} in Section~\ref{sec:mechanism}. The materials domain bound record describes a table of computed band gaps. It names the approximation method that produced the values and declares them not interchangeable with values from another method or with measured values. It also states that a gap of zero means the calculation predicts a metal, not that a measurement is missing. It leaves the onset reconstruction method blank, since a band gap has no onset.

Values in a domain bound are identifiers where a community maintains them and names where it does not. The materials record gives the approximation method as \texttt{libxc:GGA\_X\_PBE}, an entry in the functional library the major electronic structure codes already share. The sepsis record gives the operational definition as the DOI of the paper defining it~\cite{singer2016}, because clinical criteria have no comparable registry. Matching is exact, so two parties citing the same criterion through different identifiers would not agree, and reconciling identifiers is outside the scope of this paper.

We wrote three detectors, one for each kind of record a consuming workflow could consult before accepting a reuse. Every trial gives all three the same artifact and the same proposed reuse. The only difference between them is which record each one is allowed to read.
The provenance detector reads the lineage record. It checks that the artifact is registered, that a recorded activity produced it, and that the activity names its inputs and a responsible agent. The descriptive detector reads the descriptive metadata. It checks that the artifact is findable, typed, licensed, and in the format the consumer expects. The bounds detector reads the domain bound record and evaluates the constraints compiled from it.

We built the grouping by assigning each fault class in Table~\ref{tab:faults} to the record that would need to contain the information to catch it. That assignment is ours, made to separate what provenance, descriptive metadata, and domain bounds each account for. The residual classes have no detector, since no record in this study contains that information. Datasheets for Datasets and PROV-AGENT do what they were built to do, making a dataset's limitations legible to a reader and recording what a workflow did. Neither was designed to answer whether one particular proposed reuse conforms to the conditions an artifact was prepared under.

Each of the 16 fault classes was injected against both records, giving 32 faults in total, alongside 144 benign controls. A benign control is a reuse request that violates nothing, including requests that state only some of the conditions and leave the rest unstated. The controls are there so the result can come out wrong: a detector that reported a violation every time would catch all 32 faults, so the detection rate means something only if the bounds detector tells the faults and the controls apart.

\subsection{Detection}

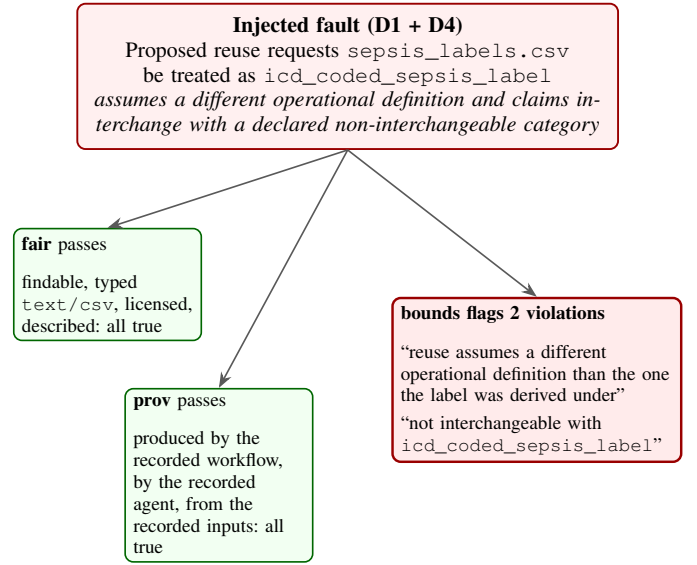
\begin{figure}[t]
\centering
\resizebox{\columnwidth}{!}{%
\begin{tikzpicture}[
  inject/.style={
    rectangle, rounded corners=3pt,
    draw=red!60!black, line width=0.7pt,
    fill=red!6,
    text width=6.6cm, align=center,
    font=\footnotesize, inner sep=5pt
  },
  pass/.style={
    rectangle, rounded corners=3pt,
    draw=green!40!black, line width=0.6pt,
    fill=green!5,
    text width=2.2cm, align=left,
    font=\scriptsize, inner sep=3pt
  },
  flag/.style={
    rectangle, rounded corners=3pt,
    draw=red!60!black, line width=0.9pt,
    fill=red!8,
    text width=3.5cm, align=left,
    font=\scriptsize, inner sep=3pt
  },
  arr/.style={-{Stealth[length=2mm]}, line width=0.6pt, draw=black!65},
]

\node[pass] (fair) {
  \textbf{fair} passes\\[6pt]
  findable, typed \texttt{text/csv}, licensed, described:
  all true
};
\node[pass, below right=6mm and -10mm of fair] (prov) {
  \textbf{prov} passes\\[6pt]
  produced by the recorded workflow, by the recorded agent,
  from the recorded inputs: all true
};

\coordinate (leftmid) at ($(fair.center)!0.5!(prov.center)$);
\node[flag, anchor=west, xshift=10mm] at (prov.east |- leftmid) (bounds) {
  \textbf{bounds} \textbf{flags 2 violations}\\[6pt]
  ``reuse assumes a different operational definition than
  the one the label was derived under''\\[2pt]
  ``not interchangeable with \texttt{icd\_coded\_sepsis\_label}''
};

\coordinate (midx) at ($(fair.west)!0.5!(bounds.east)$);
\node[inject, anchor=south, yshift=10mm] at (midx |- fair.north) (inject) {
  \textbf{Injected fault (D1 + D4)}\\
  Proposed reuse requests \texttt{sepsis\_labels.csv} be treated as
  \texttt{icd\_coded\_sepsis\_label}\\ \emph{assumes a different
  operational definition and claims interchange with a
  declared non-interchangeable category}
};

\draw[arr] (inject.south) -- (fair.north);
\draw[arr] (inject.south) -- (prov.north);
\draw[arr] (inject.south) -- (bounds.north);

\end{tikzpicture}%
}
\caption{One injected fault, seen by all three detectors. The reuse asks to treat a Sepsis-3 artifact as an ICD-coded one, combining faults D1 and D4 in a single request. Provenance and descriptive metadata both pass, since everything they check is true of the artifact; neither checks the operational definition the label was derived under, which is the condition this reuse violates.}
\label{fig:worked}
\end{figure}

Figure~\ref{fig:worked} follows a single injected fault through all three detectors. The reuse asks to treat a Sepsis-3 artifact as an ICD-coded one, combining D1 and D4 in a single request. The provenance and descriptive detectors both pass, as everything they check is true of the artifact. Neither checks the operational definition the label was derived under, which is the condition this reuse violates. The bounds detector reports it.

Table~\ref{tab:detection} reports how often each detector caught each group of faults. Each group was caught by one detector and missed by the other two. Domain-bound faults were caught in 17 of 18 trials by the bounds detector and in none by the other two. Lineage faults were caught by the provenance detector, and in half of trials by the descriptive detector. Descriptive faults were caught only by the descriptive detector, and neither group by the bounds detector. No detector reported a violation on any of the 144 controls.

%

\begin{table}[th]
\caption{How often each detector caught each group of faults. Rows are what
went wrong, columns are which detector caught it.}
\label{tab:detection}
\centering
\begin{tabular}{lrrrr}
\toprule
Fault group & $n$ & Provenance & Descriptive & Domain bounds \\
\midrule
Domain-bound & 18 & 0.00 & 0.00 & \textbf{0.94} \\
Lineage & 4 & \textbf{1.00} & 0.50 & 0.00 \\
Descriptive & 6 & 0.00 & \textbf{1.00} & 0.00 \\
Residual & 4 & 0.00 & 0.00 & 0.00 \\
\midrule
False positive rate & 144 & 0.00 & 0.00 & 0.00 \\
\bottomrule
\end{tabular}
\end{table}

The other two detectors did not catch domain-bound faults, which is expected. A system with complete lineage and complete descriptive metadata still accepts every one of these reuses and reports nothing, so a domain bound adds a record that is currently missing.

One domain-bound fault was missed. It falls on the materials record under D2, onset reconstruction mismatch, which that record leaves blank because a band gap has no onset. With nothing stated, no constraint is compiled, so there is nothing to violate. The checker can only catch violations of recorded conditions.

\subsection{A measured silent fault}

The previous experiment compares kinds of metadata record and shows that a domain bound refuses reuse requests the others cannot detect. It does not show what happens when such a reuse goes through. We measure that here, using published patient data.

Chicco and Jurman distribute two sepsis cohorts drawn from the same Norwegian hospital population between 2011 and 2012~\cite{chicco2020}. The primary cohort was assembled before the Sepsis-3 criterion, and the study cohort is the subset that meets Sepsis-3. Nothing else differs between them, so the definition alone accounts for what changes. Applying Sepsis-3 shrinks the cohort from 110,204 admissions to 19,051, raises the mortality rate from 7.4\% to 18.9\%, and raises the mean age of the patients included from 62.7 to 72.5 years.

We fit a logistic regression model to predict whether a patient dies, using the Sepsis-3 cohort, and scored it on the wider population assembled before Sepsis-3. AUROC did not fall. It was 0.589 on held-out Sepsis-3 patients and 0.705 on the wider population, which has a broader spread of ages and is easier to sort. Chicco and Jurman report the same ordering from support vector machines and gradient boosting, close to 0.7 on the primary cohort and 0.586 on the study cohort~\cite{chicco2020}, so our simpler model reproduces their study-cohort figure and their ordering. The predicted death rate, however, was wrong: in bounds the model came within about 0.1 percentage points, and out of bounds it predicted 9.2 points too many against a rate of 7.4\%. Neither run raised an error or a warning.

The model was fitted where 18.9\% of patients died, so its estimates average near that figure. Fitting does not happen again when the model is applied to new patients, and in the wider population 7.4\% died. The model therefore reports more than twice as many deaths as actually occurred. AUROC compares patients against each other and does not change when every estimate moves by a similar amount, which is why it did not fall. A team checking AUROC would see a model that had improved. A domain bound stating the Sepsis-3 population, checked against a request to use these labels for the wider one, fails on D1 and D3 before any model is fitted.

\subsection{Future work: Residual classes not detected}

Two residual classes in our fault taxonomy were uncaught. Both are identified in the literature, and are examples of targeted future work. 

A \emph{mis-authored bound} (R1) states conditions which do not apply to an artifact. A consuming workflow could read the bound, declare a reuse consistent with it, and pass validation, because the bound is the only account of the artifact the consumer has. Catching such errors means checking the bound against something other than the request. Wang et al. give methods for measuring how a concept's label, intension, and extension move between versions of a knowledge organization system~\cite{wangdrift2011}. Where a bound cites identifiers from a versioned vocabulary, it could be re-checked each time that vocabulary is revised rather than only when it is written.


The second class is an \emph{unmodeled dimension} (R2): a field with no stated constraint, which therefore cannot produce a violation. We treat such a field as unconstrained. A blank field could mean the dimension does not apply, or that someone left it out, and the record does not say which. The materials record has no onset reconstruction method because a band gap has no onset, but a sepsis record could omit the same field by oversight. Separating those two cases would have caught the single fault our study missed. Doing so in general is harder, since experts rely on shared knowledge and assumptions their data labeling does not capture~\cite{lebovitz2021}. Domain bounds surface judgments about data labels and values so they are recorded and checkable; they do not make those judgments correct.


\subsection{Discussion}

Detection happens at \emph{reuse request} time, which in practice means the choice of a dataset or of a model trained on particular datasets, and the orchestration or deployment of the workflow. Both sides of the check exist at those points, so a class of silent data fault can be eliminated \emph{before} workflow execution, at a cost that does not vary with the size of the dataset. Our experiments ran on a commodity laptop, with completion times on the order of tens of microseconds, which is compatible with checking during distributed job scheduling or workflow orchestration.

As noted in the setup, the assignment of fault classes to records is ours, so part of the separation in Table~\ref{tab:detection} is constructed rather than discovered. We limited the effect by including fault classes those records do catch, by drawing benign controls that include partial requests, and by treating an unasserted dimension as unconstrained rather than as a violation. That last choice has a consequence of its own: a consumer who declares nothing passes every check. A validator could instead report which conditions it was unable to evaluate, rather than passing silently.

The fault taxonomy is grounded in published cases but is not exhaustive, and the relative frequency of these classes in practice is unknown. We report detection rate per injected group and make no claim about how often each occurs.

The measured fault rests on three features: age, sex, and prior septic episode count. In-bounds AUROC of 0.589 reflects what those three features support rather than the choice of model, since Chicco and Jurman report 0.586 on the same cohort using support vector machines and gradient boosting~\cite{chicco2020}. The primary cohort contains the study cohort, so scoring against it includes patients the model was fitted on, and the 9.2 point calibration error is smaller than a disjoint population would show. No patient identifier is distributed with the cohorts, so this is a transfer between populations selected by two definitions rather than two labels over the same patients. Measuring the latter would need a credentialed database such as MIMIC-IV. We release cohort definitions for that measurement but have not run it.

Our validator implements the subset of SHACL Core used by the compiled constraints, rather than the full specification. The emitted shapes are ordinary SHACL and validate identically under \texttt{pyshacl} across all twenty conforming and violating cases.

\section{Conclusion}
\label{sec:conclusion}

A new data-centric ecosystem is emerging in computational science, driven by workflows that derive insights from existing datasets and from AI models trained on those datasets. Correct and reliable reuse of scientific data depends on whether the conditions under which categories, measures, and labels were stabilized are recorded. When they are not, errors travel through retrieval, integration, and model development, and the workflows depending on those components have no way to detect them. We have identified this as a silent fault class affecting computational science workflows, given it a taxonomy built from published cases, and shown that checking authored domain bounds detects it where lineage and descriptive records do not. Using two published sepsis cohorts, we measured what one such fault costs: a model ranked patients better than it had in bounds, while reporting more than twice the number of deaths that occurred. Our approach uses standard frameworks for data description and constraint checking, and detection happens before workflow execution.

Domain-bound constraint profiles should next be developed with domain communities and evaluated on whether they improve retrieval, integration, and transfer decisions in practice. Profiles should include explicit versioning, so that a bound can be re-checked automatically when a vocabulary it cites is revised. Building the check into workflow orchestration frameworks and distributed schedulers would make it automatic rather than something a consumer has to invoke. A model trained on bounded data could also be asked what conditions its training data was prepared under, which would extend the approach from datasets to the models derived from them.


\section*{Acknowledgment}

This manuscript has been authored by UT-Battelle, LLC, under contract
DE-AC05-00OR22725 with the US Department of Energy (DOE). The publisher, by
accepting the article for publication, acknowledges that the U.S.
Government retains a non-exclusive, paid up, irrevocable, worldwide license
to publish or reproduce the published form of the manuscript, or allow
others to do so, for U.S. Government purposes. The DOE will provide public
access to these results in accordance with the DOE Public Access Plan
(http://energy.gov/downloads/doe-public-access-plan).

\renewcommand{\bibfont}{\footnotesize}
\printbibliography

\end{document}